\documentclass[psf,epsf,twocolumn,showpacs,preprintnumbers,floatfix]{revtex4}
\usepackage{graphics}
\usepackage{graphicx}
\usepackage{dcolumn} 
\usepackage{bm}
\usepackage{epsfig}
\newcommand{\sto}{SrTiO$_3$} 
\newcommand{\lto}{LaTiO$_3$}
\newcommand{\lao}{LaAlO$_3$}
\newcommand{\tith}{Ti$^{3+}$}
\newcommand{\tif}{Ti$^{4+}$}

\newcommand{\eg}{{\it e.g.}}
\newcommand{\etal}{{\it et al.}}

\begin{document}
\title{
Avoiding the polarization catastrophe in LaAlO$_3$ overlayers on
    SrTiO$_3$(001) through a polar distortion
}
\author{Rossitza Pentcheva$^{a}$ and Warren E. Pickett$^{b}$}
\affiliation{$^{a}$Department of Earth and Environmental Sciences, 
University of Munich, Theresienstr. 41, 80333 Munich, Germany}
\affiliation{$^b$Department of Physics, 
  University of California, Davis, One Shields Avenue, Davis, CA 95616, U.S.A.}
\date{\today}
\begin{abstract}
A pronounced uniform polar distortion extending over several unit cells 
enables
thin LaAlO$_3$ overlayers on SrTiO$_3$(001) to counteract the charge dipole and 
thereby neutralize the ``polarization catastrophe'' that is suggested by simple 
ion-counting.  
This unanticipated mechanism, obtained from density 
functional theory calculations, allows several unit cells of the  LaAlO$_3$ overlayer
to remain insulating (hence, fully ionic). The band gap of the system, 
defined by occupied O $2p$ states at the surface and
unoccupied Ti $3d$ states at the interface in some cases  $\sim$20~\AA~distant,
decreases with increasing thickness of the LaAlO$_3$-film before an
insulator-to-metal transition and a crossover to an electronic reconstruction  
occurs at around five monolayers of LaAlO$_3$.
\end{abstract}
\maketitle

Heterostructures containing polar (\eg\ \lao\ (LAO)) and nonpolar (\eg\ \sto\ (STO)) oxides
raise issues similar to semiconductor heterostructures, \eg~Ge/GaAs~\cite{harrison78}. If 
 the ionic charges are maintained, then adding LAO layers (with alternating positively 
charged  LaO and negatively charged  AlO$_2$-layers) onto an STO 
substrate is expected to lead to an ever-increasing dipole. The ´´accompanying potential shift across the LAO
slab should eventually cause a ``polarization catastrophe''\cite{nakagawa06,noguera08}. 
To avoid this, surfaces are supposed to reconstruct (atomically, \eg\ via defects) 
or facet~\cite{tasker}. However, in oxides containing transition metal ions further compensation mechanisms 
are available. These can lead to electronic phenomena that are
unanticipated from the properties of the bulk constituents and can thus invoke 
new functionality. Although LAO
and STO are two 
conventional nonmagnetic band insulators, a high  
conductivity~\cite{ohtomo2004} as well as indications for magnetism~\cite{brinkman07} and 
superconductivity~\cite{reyren07} were measured at the $n$-type LaO/TiO$_2$ interface (IF). 
The initially reported high carrier density was subsequently found to be sensitive
to film growth conditions~\cite{nakagawa06,brinkman07,siemons07,kalabukhov07,basletic08,willmott07}. Suppressing oxygen vacancies leads to an increase of the sheet resistance by several orders of magnitude ~\cite{brinkman07,rijnders08}.

Recently, a thickness dependent switching between insulating and conducting behavior~\cite{thiel06} 
was reported around four monolayers (MLs) of LAO on STO(001). Thiel \etal~\cite{thiel06}
found that the insulator-to-metal transition can also be induced dynamically by an external electric field.
Writing and erasing of nanoscale conducting regions in LAO overlayers on 
\sto(001) was demonstrated~\cite{cen08} with an applied local electric field (by atomic force microscope). 
In contrast to the abrupt transition in LAO films on STO(001), Huijben and collaborators~\cite{huijben06} 
observed a smoother increase in carrier density in coupled $p$- (hole-doped) and $n$-type (electron-doped) interfaces 
where the LAO film was covered by \sto.   
These results lend additional urgency to  understand the
insulator-to-metal switching in oxide nanostructures.

The seeming violation of charge neutrality has become a central consideration
in trying to understand the unexpected behavior that emerges at the interface. 
So far, theoretical efforts have concentrated mainly on  LAO/STO 
superlattices~\cite{pentcheva06,seifert06,park06,albina07,maurice07,janicka08,zhong08,pentcheva08}. 
For these, density functional theory  calculations (DFT) suggest that the charge 
mismatch at the electron 
doped $n$-type interface is accommodated by partial occupation of the Ti $3d$ band.  At low temperatures a 
disproportionation on the Ti sublattice 
in the interface layer is predicted within LDA(GGA)+U~\cite{anisimov93} with \tith~and \tif~arranged in a checkerboard 
manner~\cite{pentcheva06,zhong08,pentcheva08}. 

At first glance the transition from insulating to conducting behavior found between three and four  
MLs on \sto(001)~\cite{thiel06} does not fit into the 
picture obtained from the GGA+U calculations for infinitely extended 
superlattices~\cite{pentcheva06,pentcheva08}. In the latter the charge mismatch is accommodated
immediately at the IF and largely independent of the thickness of LAO and STO. 

Here we shed light on the nature of the electronic state at the interface 
and the origin of thickness dependent metal-to-insulator transition. Based on DFT-calculations, we demonstrate
 that an interface between a thin LAO film on a 
STO(001) substrate is fundamentally different from an IF in a 
periodic superlattice. In a thin overlayer 
there are two polar discontinuities, one   at 
the IF and another  at the surface.  
As we will show in this paper, the 
proximity to the surface and the associated atomic relaxation
has a profound effect on the properties and 
compensation mechanism at the interface. 

To address these issues, we study 
here the structural and electronic properties of 1-5 ML 
of LAO on a STO(001)-substrate, and contrast them with the corresponding
behavior at LAO/STO superlattices. The DFT calculations are performed with the all-electron full-potential augmented plane 
waves (FP-APW) method in the \textsf{WIEN2k}-implementation~\cite{wien} and  
generalized gradient approximation (GGA)~\cite{pbe96} of the exchange-correlation 
potential. 
We have used the lateral lattice constant of the \sto-substrate obtained 
from GGA (3.92\AA) because it is in 
closer agreement with the experimental value (3.905\AA). 
Thus the  epitaxial
\lao-film (3.79\AA) is subject to tensile strain on \sto\  
due to a lattice mismatch of 3\%. 
The GGA-gap for bulk \sto\ is 2.0 eV (experimental value 3.2 eV) separating 
filled O $2p$ bands and unfilled Ti $3d$ bands. For \lao\ we obtain a band gap 
within GGA of 3.7~eV (experimental value 5.6 eV) between filled O $2p$ bands 
hybridized with 
Al $p$-bands and unfilled conduction bands comprised of La $5d$  and Al $3s, 3p$ 
states.

The thin (1-5 ML) LAO films on a STO(001) substrate were modeled in the supercell geometry with two inversion 
symmetric surfaces to avoid spurious electric fields. 
To minimize the interaction between neighboring surfaces, the slabs are separated  in $z$-direction 
by 10-12\AA\ of vacuum. Because XRD~\cite{vonk} and LEED~\cite{moritz} measurements give no 
indication for a superstructure, we have considered only defect-free unreconstructed surfaces.   
However, we have allowed for more complex symmetry breaking and electronic reconstruction by 
using a $c(2\times 2)$-lateral periodicity. For this cell 15 
$k$-points in the irreducible part of the Brillouin zone were used.


\begin{figure}[t]
    \begin{center}
 \includegraphics[scale=0.4]{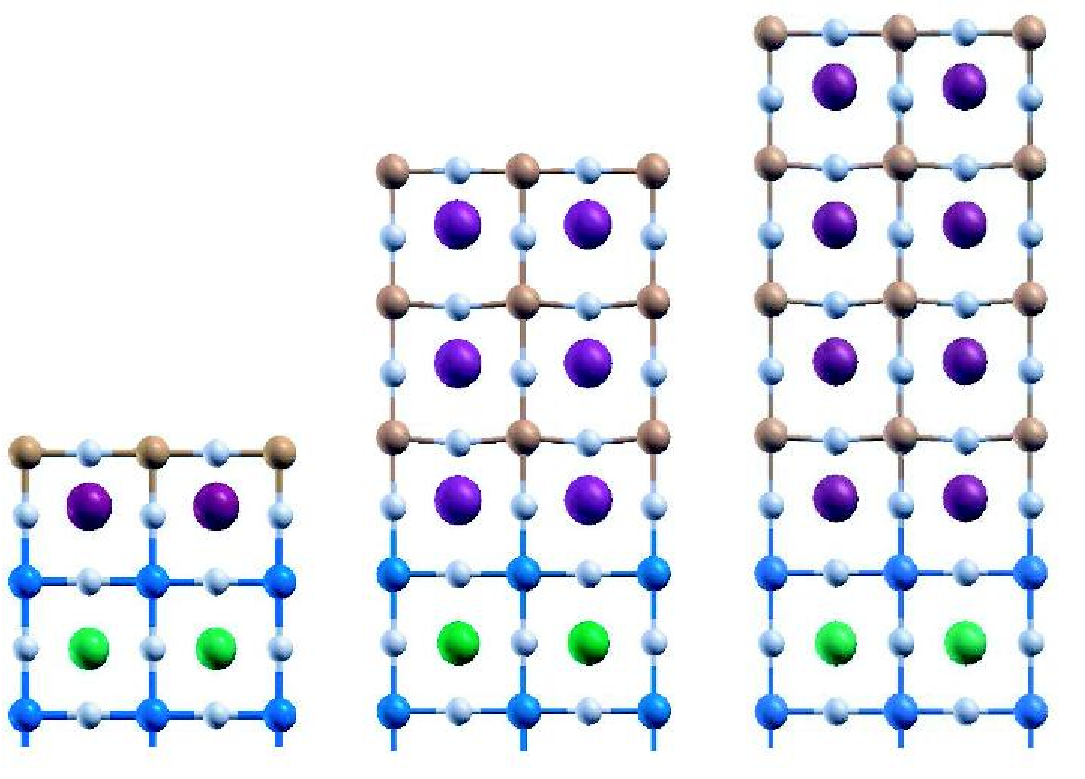}
    \end{center}
\caption{\label{fig:surfstr} Side view of the relaxed structures of 1, 3 and 4 ML LAO on STO(001) showing the ferroelectric distortion. The oxygen ions are marked by light grey spheres, while the Sr-, Ti-, La- and Al-cations are shown by green, blue, purple and orange spheres. }
\end{figure}
\begin{figure}[t]
    \scalebox{0.7}
  {\includegraphics{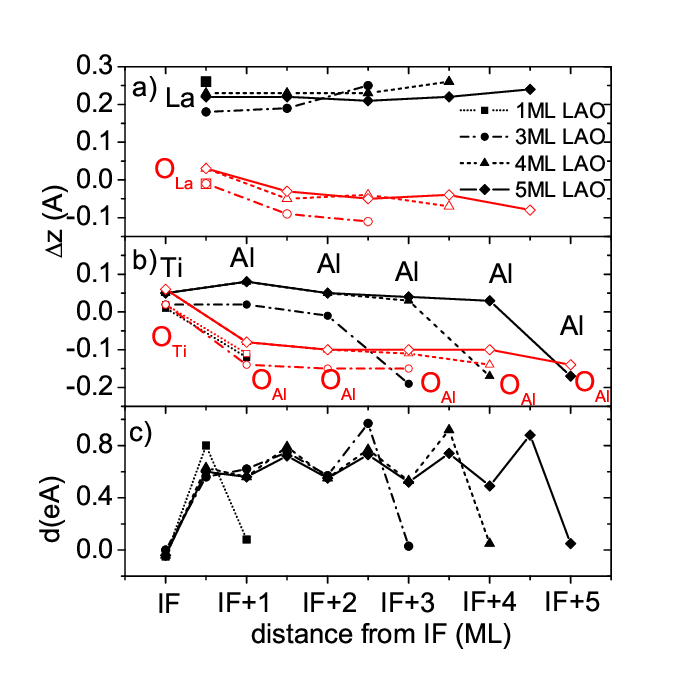}}
\caption{\label{fig:dzLAOSTO001}
Vertical displacements of ions $\Delta z$ with respect to the bulk positions  in \AA\ for 1-5 MLs of LAO on STO(001)  obtained within GGA.   a) and b) displays the relaxations in the AO and BO$_2$ layers, respectively with cation/oxygen relaxations marked by full black/open red symbols. The corresponding layer-resolved dipole moments are displayed in (c). The $x$ axis shows the distance from the interface (I) TiO$_2$-layer.}
\end{figure}

{\it Polar distortions}.
A full structural optimization was performed within the tetragonal unit cell using GGA. The relaxed structures are 
pictured in Fig.~\ref{fig:surfstr} and the atomic displacements 
which are exclusively along the $c$-axis are plotted in Fig.~\ref{fig:dzLAOSTO001}. 

An unexpected pattern is uncovered: $(i)$ The 
surface AlO$_2$ layer shows an inward relaxation of both Al (0.12-0.19\AA) and O (0.11-14\AA) atoms by
similar amounts, resulting in only a small buckling of 0.01-0.04\AA. $(ii)$ The most striking result 
is a strong buckling  
in all LaO layers with a uniform polar-type distortion of 0.20-0.36 \AA.
This relaxation is dominated by the outward movement of La by up to
0.26 \AA. $(iii)$ The subsurface 
AlO$_2$ layers are also buckled by approx. 0.15 \AA: Al relaxes outward by 0.01-0.08~\AA,  
O inward by $\sim0.10$~\AA.
$(iv)$ The Ti and O-ions at the 
interface show only small outward relaxation (less than 0.06~\AA). 
This situation of large polar distortions throughout the LAO slab with
negligible distortion in STO contrasts strongly the behavior obtained
for \lao\ (or \lto)/\sto\ superlattices
where ferroelectric-like plane-buckling is predominantly in the TiO$_2$ layer at the 
interface~\cite{hamann06,spaldin06,seifert06,park06,albina07,maurice07} and  
amounted to 0.17\AA\ for a 5LAO/5STO heterostructure~\cite{pentcheva08}. 

Although not explicitly discussed, previous VASP calculations showed a similar relaxation 
pattern~\cite{cen08}. In contrast, another WIEN2k study reported  only small 
relaxations~\cite{schuster08} also finding metallic behavior for all LAO thicknesses. 
Vonk {\it et al.}\cite{vonk} fitted structural models to surface x-ray
diffraction (SXRD) data  for a LAO overlayer on STO(001) with a total coverage 
of 0.5 MLs. The best fit infers 
a contracted TiO$_6$ octahedron at the interface and an outward relaxation of La in the LaO layer. 
A direct comparison to our results is not possible because of the different coverage and 
since the incomplete LAO layer 
allows for lateral relaxations of strain.

{\it Compensating electronic and ionic dipoles}.
The implied dipole moment of a 4ML \lao-film is  
${\cal D}_{bare} \approx -0.5 e \times (3.9 \AA) \times 4 {\rm cells}  =
 -7.8~e\AA $ 
which translates 
into a bare potential shift across the four LAO layer slab
of 80-90 eV.  This electric field is screened by electronic {\it plus} ionic
polarization.  Using the bulk LAO dielectric constant $\epsilon$=24, this
becomes approximately
$\Delta V = 4\pi e \frac{{\cal D}_{bare}}{\epsilon (3.9 \AA)^2} \sim 3.5~{\rm eV},$
which can be sustained by the 5.6 eV gap of LAO.

The dipole shift due to the polar distortion can be estimated from the
displacements, using the formal ionic charges~\cite{Born}. 
The layer resolved dipoles, pictured in 
Fig.~\ref{fig:dzLAOSTO001}, show a strikingly uniform behavior in the inner LAO layers (a strong 
dipole moment of $\sim 0.55 $ and $\sim 0.75$ $e$\AA\ in the AlO$_2$ and LaO layers, 
respectively) and an almost vanishing dipole in the interface TiO$_2$ and surface AlO$_2$ layers.  
The total ionic dipole \eg\ for the 4ML LAO film is: 
\begin{eqnarray}
{\cal D} = \sum Z_i \Delta z_i \approx
{\rm 4.8~e\AA} 
\end{eqnarray}
This rough estimate gives a value that is more than 60\% of ${\cal D}_{bare}$ and of opposite sign.  
The large polar distortion has a significant impact on the electronic behavior of the system (LAO film + STO substrate).


{\it Electronic properties of overlayer + interface.}
\begin{figure}[t]
    \begin{center}
\includegraphics[scale=1.1]{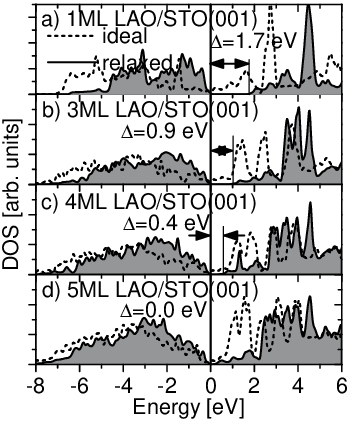}
    \end{center}
\caption{\label{fig:surfdos} Density of states for the ideal (dashed line) and 
relaxed (solid line, grey filling) structure of 1-5 ML LAO on STO(001).
Relaxation opens a band gap, but its size decreses with each added LAO layer.}
\end{figure}
\begin{figure}[t]
    \begin{center}
 \includegraphics[scale=1.]{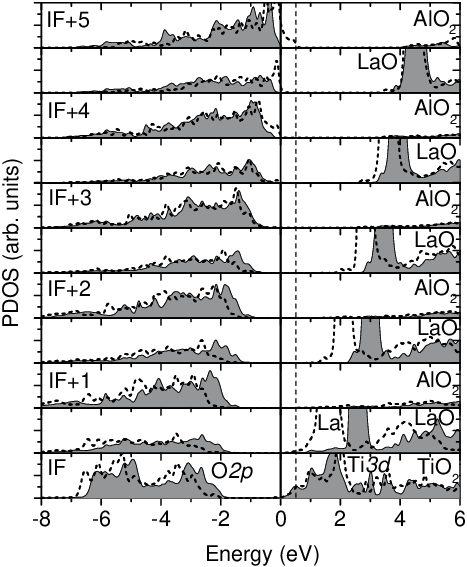}
    \end{center}
\caption{\label{fig:surfdosOp} Layer resolved density of states of 5 ML LAO on STO(001) with ideal (dashed line ) and relaxed (grey shaded area) coordinates. The DOS for the ideal positions was shifted by 0.5 eV to align with the conduction band of the system with the relaxed atomic positions. The Fermi level of the unrelaxed system 
is marked with a dashed line.
}
\end{figure}
The total density of states as a function of the LAO thickness for 1-5 ML thick LAO films
on STO(001) is shown in Fig.~\ref{fig:surfdos}. For the ideal atomic positions 
all systems are metallic due to an overlap of the surface O $2p$ bands with the Ti $3d$ 
bands at the interface (cf. Fig. \ref{fig:surfdosOp}). The strong lattice polarization leads to insulating 
behavior for $N=1-4$ ML. With increasing thickness of the LAO-film the gap
decreases by approximately 0.4 eV per added LAO-layer starting from 1.7~eV for 1ML LAO 
and finally closes for 5 ML LAO/STO(001). While the DFT calculations correctly predict 
an insulator-to-metal transition, the critical thickness is affected by the band 
gap underestimation within GGA and is expected to occur beyond an LAO thickness of 
6 MLs for a defect-free surface. Defects and/or adsorbates which are not considered in the 
theoretical model may be responsible for the lower critical thickness found in experiment~\cite{thiel06}. 

In order to gain more insight into the type of electronic states around 
the Fermi level and their spatial distribution, we have plotted in 
Fig. \ref{fig:surfdosOp} the layer-resolved density of states for 
5ML LAO/STO(001) with ideal and relaxed atomic positions.  A nearly rigid 
upward shift of the O $2p$-bands towards the Fermi level is observed as they approach the surface.  
For the relaxed configuration 
the top of the valence band is determined by the O $2p$ band in the surface AlO$_2$ 
layer, while the bottom of the conduction band is fixed by the unoccupied Ti $3d$ band at the interface.
Thus, the closing of the gap is due to the raising of the O $2p$ valence band
states in the surface LAO layer as the LAO slab gets thicker.

For an isolated IF (in LAO/STO superlattices),
the ``extra'' $0.5e^-$ per IF Ti ion must be accommodated,
it cannot go into the LAO layers due to the large band gap, and it
remains at the IF and charge-orders (within GGA+U) rather than occupy 
itinerant Ti $3d$ states within the STO layers~\cite{pentcheva06,pentcheva08}.  In contrast, in thin LAO
overlayers, it does not go into the Ti $3d$ states {\it anywhere}; the lattice screening preserves 
the formal electronic 
charges corresponding to closed shells, therefore including correlation effects within LDA+U would
 not influence the result.  Up to a critical LAO thickness of 5ML 
the `extra' 0.5 electron at the interface appears to have been shifted to
the surface and there is no need for an electronic (or atomic) reconstruction 
as observed for the superlattice. However, beyond a critical coverage, the system is expected to 
become metallic.


In summary, strong lattice polarization compensates the dipolar electric field in thin LAO-overlayers 
and sustains the insulating behavior up to 5 MLs LAO. The effect of this polar distortion 
is crucial: Without it, the system is metallic for any coverage of LaAlO$_3$. The band gap 
in the relaxed systems formed between the O$2p$ states in the surface layer and the Ti $3d$ states at the 
interface decreases gradually and collapses at 5 ML LAO/STO(001) giving rise to 
an electronic reconstruction as the system approaches the limit of the isolated interface. 

The structural distortion we observe here differs from issues discussed for ultrathin ferroelectric  
films, \eg\ the existence of a critical thickness below which ferroelectric displacements are 
suppressed~\cite{Fong04,Dawber,Despont,FErev}. 
In contrast, \lao\ is a wide band gap insulator with a low dielectric 
constant and no ferroelectric instability. In this case the lattice distortion is driven by the polar 
nature of the surface. Further examples that nanoscale objects can sustain polarity are 
recently emerging (see \cite{Goniakowski} and references therein). In the context of tailoring 
materials properties for device applications, finite size effects in ultrathin oxide films 
prove to be an exciting area for future research.

We acknowledge useful discussions with H.Y. Hwang, M. Huijben, J. Mannhart, N. Spaldin and 
M. Stengel and support through the Bavaria-California Technology Center 
(BaCaTeC), ICMR,  DOE Grant DE-FG03-01ER45876, 
and a grant for computational time at the Leibniz Rechenzentrum.


\begin{thebibliography}{10}
\bibitem{Born}We note that both the shift by a rigid number of lattice constants and the use of 
formal charges (or Born effective charges~\cite{seifert06} which may be more approapriate) is only a rough estimate.
\end{thebibliography}
\end{document}